# Design and Analysis of a multiple quantum-well plasmonic Germanium/Silicon-Germanium nanolaser

Hamed Ghodsi [1] • Hassan Kaatuzian [1]

**Abstract** There are many researches reported in using Germanium in Silicon based lasers but acquiring this potential for a nanolaser may also be important for development of a CMOS compatible plasmon source. In this paper, a Ge/SiGe multiple quantum well waveguide integrated nanolaser is introduced and theoretically investigated. This structure is simulated and by means of a semi-classical rate equation model, its performance is studied. The proposed nano laser has a tiny footprint of 0.07µm$^2$, room temperature performance and CMOS compatible fabrication process. The output performance of the proposed structure as estimated, is noticeable. These simulated results, compared with some experiments and redundant software double checking, show acceptable compatibility. In 1550nm output wavelength, it provides 3.83µW output power with 1µA injection current while maintaining its performance in a wide modulation bandwidth of 24.5GHz. This remarkable performance is achieved thanked to a Purcell factor equal to 2208.

Hamed Ghodsi
hamed_88@aut.ac.ir

Hassan Kaatuzian
hsnkato@aut.ac.ir

[1] Photonics Research Laboratory (PRL), Electrical Engineering Department, Amirkabir University of Technology

## 1. Introduction

Plasmonic integrated devices and circuits are considered as the future of integrated photonics and optoelectronic devices and systems due to their nano dimensions and tremendous bandwidth. [1,2,3,4] However, despite several number of researches on design and fabrication of plasmon sources i.e. plasmon nanolasers or SPASERs which can be categorized in metallic nanoshells [5], nanocavities [6], nanowires [7] and waveguide-based nanolasers [8], there is almost no plasmon source suitable for introducing commercially available plasmonic integrated circuits, compatible with CMOS fabrication process. There are many laboratory realizations, which in many cases have optical pumping or just operate in cryogenic conditions; however, they are not ready for this purpose yet. Therefore, there is still demand for CMOS friendly silicon based plasmonic nanolasers.

On the other hand, in recent years there are several researches on transforming Germanium to a direct energy gap material for implementation of Ge/SiGe semiconductor lasers [9], which are definitely more compatible with commercial silicon based electronics than their III-V counterparts. Thus, using this potential in development of subwavelength Ge/SiGe plasmonic nanolasers will play an important role in the future of plasmonic nanolasers and making their important applications like, next generation high-speed integrated circuits, nano scale and low power light sources and medical devices [1, 10] and etc. more realistic.

In this paper, an electrically pumped multiple quantum well germanium plasmonic nanolaser is introduced and simulated. The proposed nano laser has a cubic nano resonator integrated into an insulator/metal/insulator (IMI) plasmonic waveguide. This structure while having a tiny footprint provides nearly perfect coupling to the plasmonic waveguide and so the other integrated plasmonic devices on the chip. In addition, the proposed nanolaser has the output freespace wavelength of 1.55 µm, which means it is compatible with commercial photonic devices and systems. A similar approach can be found in [6]. Our device is expected to perform normally in the room temperature and with significantly less pump current than a micro-cavity laser due to its considerably smaller footprint and thus increased level of current density with a same input current.

First, we have introduced physical structure and fabrication possibility of such a device in section 2. Then, in section 3, governing principles are explained and simulated. Following this, in section 4, results of output characteristics analysis can be seen and this paper is concluded in the final section.

## 2. Physical structure and Fabrication

3D schematics of the proposed device is sketched in Fig.1, which consists of plasmonic cube resonator placed on top of a metallic strip deposited on a SiO2 bed. In order to have better propagation length and decreasing metallic loss [1] Gold is used in the waveguide structure. Nevertheless, for better compatibility with traditional CMOS process Copper also can take place with keeping its more Ohmic loss in mind.

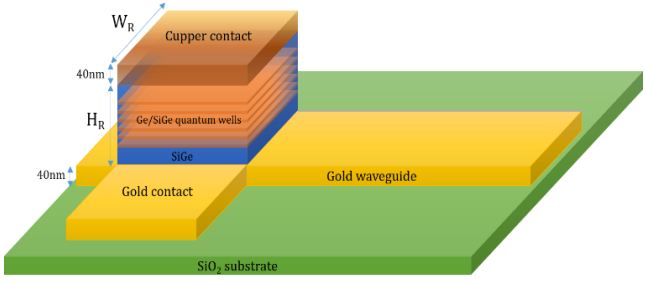

**Fig.1** 3D schematics of proposed nanolaser

Structural details about size, number of quantum wells (QWs), doping and alloy percent can be found in Table.1 and there is a Copper metal cap on the top as the second electrical contact. In addition, a lateral cross section also can be witnessed in Fig.2.

Fabrication process of such a device can be done as follows. The first step is deposition of bottom metallic strip on a $SiO_2$ substrate then it will be followed by deposition of 5nm thick $Si_{0.11}Ge_{0.89}$ layer which can be deposited by CMOS compatible process of [11] and in next step, we have three 7 nm thick quantum well Germanium layers and two 10 nm thick SiGe barriers. After that, we have to deposit a 5nm top SiGe buffer layer an eventually do the top metal contact deposition. This process as explained in [11] with 11% Silicon alloy ratio in SiGe barriers will result in 0.25% in-plane tensile strain in Germanium QWs. Such a strained layer with the aim of a relatively high donor doping equal to $7.6×10^{19}$ cm$^{-3}$ will make the Germanium, a direct bandgap material in order to have enough number of excitons for efficient plasmon generation. The calculations and theories can be found in the next sections. Finally, the process should be finished by deposition of $SiO_2$ isolation layer, which is necessary in CMOS process for deposition of top metal layers.

**Table 1.** Design parameters of the proposed nanolaser

| Symbol | Description | Value | Unit |
|---|---|---|---|
| $W_R$ | Resonator size | 265 | nm |
| $H_R$ | Resonator height | 51 | nm |
| $X_{Au}$ | Bottom metal thickness | 40 | nm |
| $X_{Cu}$ | Top metal thickness | 40 | nm |
| $X_{Bottom}$ | Bottom buffer thickness | 5 | nm |
| $X_{Top}$ | Bottom buffer thickness | 5 | nm |
| $N_{QW}$ | Number of QWs | 3 | - |
| $X_{QW}$ | QW thickness | 7 | nm |
| $X_{Barrier}$ | Barrier wall thickness | 10 | nm |
| $x$ | Ge Alloy percent | 89 | % |
| $N_D$ | Doping concentration | $7.6×10^{19}$ | cm$^{-3}$ |

## 3. Operation principles

Same as a traditional laser in a plasmonic nanolaser we should have gain medium and a resonator. First, we will discuss about resonator and mirrors and then we will explain the gain medium equations. Then we will be able to write the rate equations for simulation of the output power and pumping threshold and modulation bandwidth.

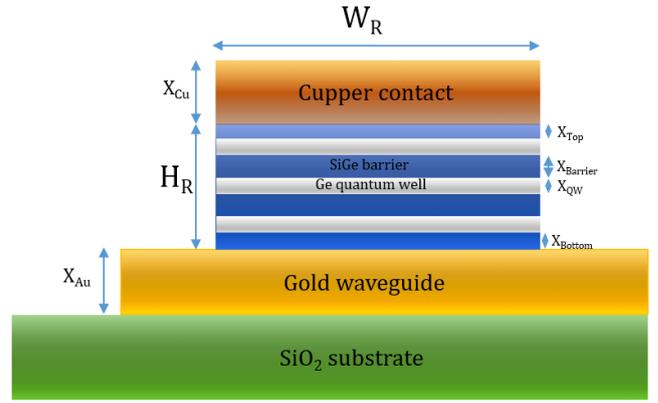

**Fig. 2** 2D Lateral cross section of the device

### 3.1. Nanocavity characterization

The plasmonic nanocavity can be characterized by resonant wavelength, quality factor, equivalent modal volume, Purcell factor and coupling factor which is also known as Beta factor.

Resonant wavelength can be calculated by finding propagation modes decay rate versus frequency in the resonator. To do so, Lumerical FDTD software package [12] is used and result is depicted in Fig.3 for cavity height equal to 51nm and for different cavity widths. It is worth mentioning that the complex dielectric constants are taken from CRC model of [12] for Gold, Copper and Germanium and taken from calculations of [13] for SiGe with different alloy ratios. In addition, effect of doping on frequency behavior of dielectric constants is neglected for simplicity.

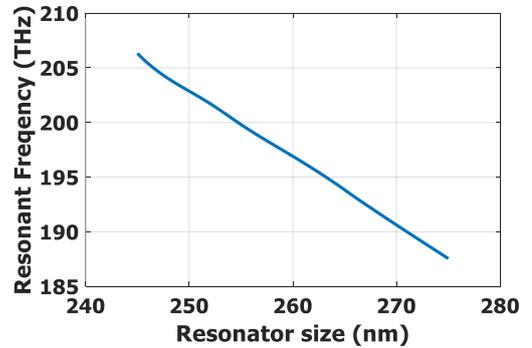

**Fig.3** Resonant frequency versus resonator size with other parameters set from Table.1

Quality factor of a plasmonic resonator can be calculated by the (1) [14]:

$$Q = 2\pi \frac{\text{Energy stored in cavity}}{\text{Energy lost per cycle to walls}} \quad (1)$$

Using design values of Table.1 and FDTD method, the Q factor is calculated and depicted in Fig.4 where the value for 1550 nm wavelength is about 12.08 which for a specific mode, it is independent of amplitude. In plasmonic metallic cavities, considering their large amount of loss, Quality factor is far less than its insulator optical counterparts are.

Effective mode volume has a key role in nanolaser operation, which can be calculated by (2). [14]

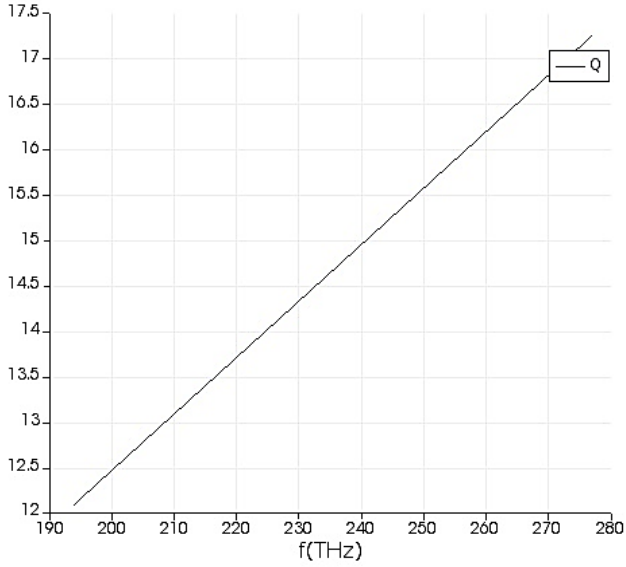

**Fig.4** Quality factor versus frequency with other parameters set from Table.1

$$V_{eff} = \frac{\int_V \varepsilon(r)|E(r)|^2 d^3r}{\max\left[\varepsilon(r)|E(r)|^2\right]} \quad (2)$$

Where "ε" is dielectric constant, "E" is the electric field and "V" is the resonator volume. From the FDTD analysis using lumerical package for the optimal values of Table.1 equivalent mode volume for different wavelengths can be seen in Fig.5 and for 193.54 THz, which is equal to 1550nm free space wavelength equivalent mode volume is about 2164 nm$^3$.

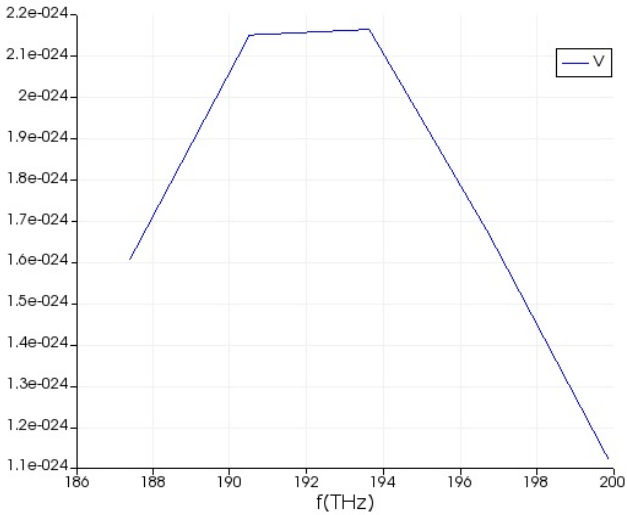

**Fig.5** Equivalent mode volume (in m$^3$) versus frequency with other parameters set from Table.1

The Purcell factor [15] "$F_p$", is a key parameter in cavity quantum electrodynamics (CQED) that defines the coupling rate between a dipolar emitter (QWs in our case) and a cavity mode. Purcell factor as can be expressed as (3) specifies the possible strategies to enhance and control light-matter interaction. [16] Efficient light-matter interaction is achieved by means of either high quality factor (Q) or low modal volume V, which is the basis of plasmonic cavity electrodynamics (PCQED).[17]

$$F_p = \frac{3}{4\pi^2}\left(\frac{\lambda}{n}\right)^3\left(\frac{Q}{V_{eff}}\right) \quad (3)$$

Where λ is free space wavelength, n is the refractive index of gain medium and Q is the quality factor of the plasmonic resonator. There is also an alternative way for finding the Purcell factor. To do so, we have to use a dipole source near the interface of bottom metal strip in the FDTD simulations and Purcell factor is equivalent to the ratio of the power emitted by a dipole source in the environment by the power emitted by the dipole in a homogeneous environment (bulk material) since the emission rate is proportional to the local density of optical states (LDOS), and the LDOS is proportional to the power emitted by the source. [12] Purcell factor for different frequencies can be witnessed in Fig.6 and the value for the output frequency is about 2208. It should be mentioned that these two approaches for calculation of Purcell factor come up with nearly the same results and we have used this fact for double-checking the calculations.

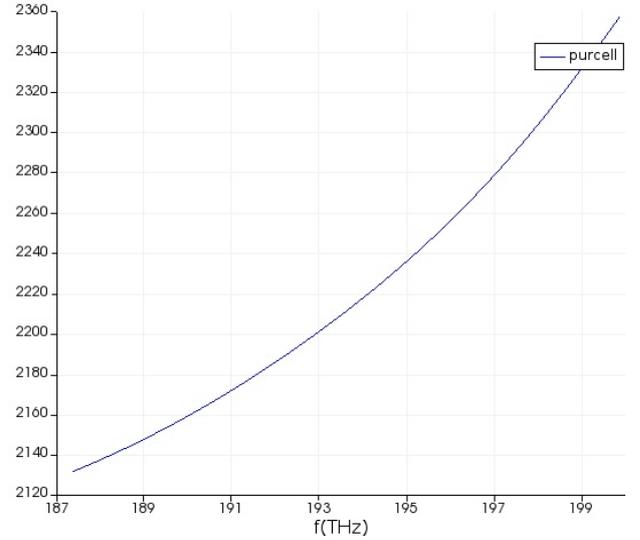

**Fig.6** Purcell factor versus frequency with other parameters set from Table.1

β which is known as coupling factor is defined by the ratio of the spontaneous emission rate into the lasing mode and the spontaneous emission rate into all other modes and can be expressed by (4).[6]

$$\beta = \frac{F_{cav}^{(1)}}{\sum_k F_{cav}^{(k)}} \quad (4)$$

Where $F_{cav}^{(k)}$ is the Purcell factor of k'th mode. k = 1 corresponds to the lasing mode and the summation is on both cavity modes and radiating modes. For calculating β factor, using lumerical FDTD package, a method based on several randomly positioned dipole sources is used where the lasing mode is determined by the dipole source with maximum Purcell factor. By means of (4) and calculation of Purcell factors for all of these dipole sources (As shown in Fig.7), β factor is determined to be about 0.324 for the proposed square cavity structure.

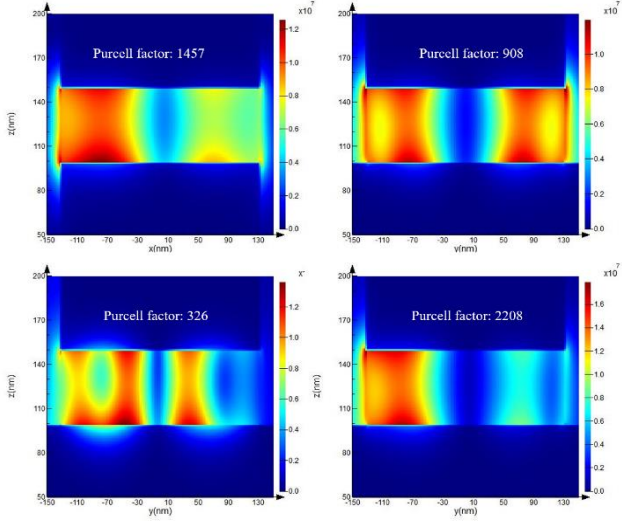

**Fig.7** Purcell factor of some of resonator modes.

### 3.2 Rate equation and output characteristics

In the proposed nanolaser structure as can be seen in Fig.8, energy of generated excitons in quantum wells due to electrical current will be transferred to Surface Plasmon Polariton (SPP) modes at the top and bottom metal semiconductor interfaces. However, the generated SPPs in the top metal contact cannot flow into the host waveguide because the top metal contact is not continued further. The reason behind this can be explained in terms of generating a massive capacitor parasitic capacitor, which will drastically degrade the modulation bandwidth of the proposed plasmon source.

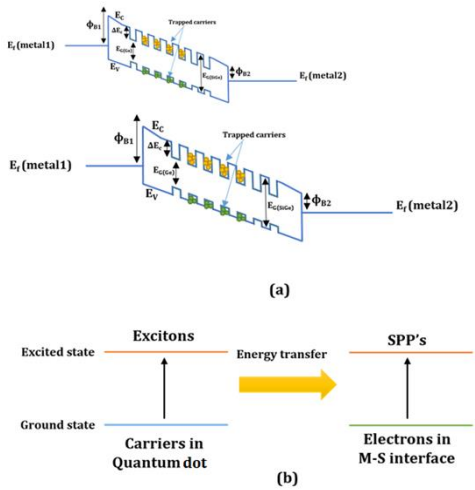

**Fig.8** Energy transfer diagram. a. Energy band diagram of nanolaser, b. Energy transfer concept

In order to analyze performance of a plasmonic nanolaser, we need a model for its rate equations. For this purpose, we will start with semi-classical rate equations proposed in [6], which can be witnessed in (5)

$$\frac{dn}{dt} = P - An - \beta\Gamma AS(n-n_0) - \frac{nv_s S_a}{V_a}$$
$$\frac{dS}{dt} = \beta An + \beta\Gamma AS(n-n_0) - \gamma S \quad (5)$$

In these equations "n" is the excited state population density of carriers in "#/cm$^3$", "S" is plasmon number in the lasing mode in "#/cm$^3$" and "P" is the carrier generation rate in the gain medium and can be roughly related to generation rate in quantum wells as expressed in (6):

$$P = P_1 + P_2 + ... + P_n \quad (6)$$

Where $P_i$'s are generation rates in the i'th quantum well. Total carrier generation rate (P) is determined by several parameters like, pump current (Injected current by electrical pumping), thermionic emission over and tunneling rates through Schottky barrier (in this case neglected due to very high doping), metal to QWs transit time (drift/diffusion theory [18]), transition probability from each well, carrier trap time in quantum wells ( indicates the average time before an exciton transfers its energy to the SPP lasing mode or lose energy due to other processes) which can be expressed by radiative recombination rate and non-radiative recombination rates [19] (Auger and SRH process) and tunneling probability between two neighbor quantum wells. All of the mentioned phenomena should be considered to achieve a precise model for finding pump rate (P) as a function of pump current (a complicated carrier dynamics model). However, in this paper we have used a simplified process, which is based only on direct and indirect recombination rates and transition probability (both tunneling and thermionic processes) between two neighbor quantum wells. [19] In this model, internal quantum efficiency (η) can be calculated from (7) which relates pump current ($I_P$) and carrier generation rate (P) as follows.

$$R_{nonradiative} = R_{Auger} + R_{SRH}$$
$$\eta = \left(\frac{R_{radiative}}{R_{radiative} + R_{nonradiative}}\right)$$
$$P = \eta \frac{I_{pump}}{q}\left(P_{QW_1} + (1-P_{QW_1})P_{QW_2} + (1-P_{QW_1})(1-P_{QW_2})P_{QW_3}\right)$$
$$P_{QW_1} \simeq P_{QW_2} \simeq P_{QW_3} = x = 1 - P_{tunnel} - P_{thermionic}$$
$$x = 1 - \exp\left(-\frac{2}{\hbar}X_{QW}\sqrt{2m_e^*(E_{G,SiGe}-E_{G,Ge})}\right) - \exp\left(\frac{(E_{G,SiGe}-E_{G,Ge})}{kT}\right)$$
$$\rightarrow P = \eta\frac{I_{pump}}{q}(3x - 3x^2 + x^3) \quad (7)$$

Where "$R_{radiative}$" is radiative recombination coefficient, which equals to "1.3×10$^{-10}$ cm$^3$/s" for Ge quantum wells of the structure [20] and "R$_{nonradiative}$" is non-radiative recombination coefficient respectively and it is equivalent to "2.285×10$^{-12}$ cm$^3$/s" which is the summation of Auger and SRH coefficient [20]. In addition, P$_{tunnel}$ and P$_{thermionic}$ are thermionic emission and tunneling probabilities between two neighbor QWs respectively. Energy bandgaps of well and barrier which are expressed by "E$_{G,Ge}$" and "E$_{G,SiGe}$" respectively can be extracted from experimental data of [11] for "L" valley of strained Germanium and SiGe layers of our structure. Eventually internal quantum efficiency can be calculated and equals to 0.9785 and Carrier generation rate is related to pump current in mA by a coefficient of 6.1156×10$^{21}$ in our case.

It is worth mentioning, that the conversion efficiency in exciton/SPP energy transfer process in the effective depth of plasmonic modes is considered 100% as a practical approximation. [7]

"A" is the spontaneous emission rate, which can be modified by the Purcell effect via "A = F$_p$A$_0$", where "A$_0$" is the natural spontaneous emission rate of the material equals to 1/τ$_{sp0}$ and τ$_{sp0}$ is the spontaneous emission lifetime of the gain medium which is Germanium QWs in our case and it is equivalent to 100ns. [20]

"Γ" which equals to the ratio of carriers generated in the spatial distribution of plasmonic modes to the whole number of generated carriers, is also called mode overlap with the gain medium coefficient and regarding mode profiles of Fig.7 it is nearly equivalent to "1". "n$_0$" is the excited state population of carriers at transparency, which is equivalent to 3.5×10$^{18}$ cm$^{-3}$ [20]. "v$_s$" is surface recombination velocity at the sidewalls of the resonator, which equals to 2160 cm/s [21]. "S$_a$" and "V$_a$" are the area of sidewalls of the nanolaser and volume of gain medium, which are equivalent to 1.325×10$^{-10}$ cm$^2$ and 3.5113×10$^{-15}$ cm$^3$ respectively. Eventually, "γ" is loss rate of plasmons per unit volume of the cavity (loss coefficient per unit length × modal speed/mode volume), which is calculated by $\gamma = \gamma_c + \gamma_g$.

"γ$_c$" and "γ$_g$" are resonator mirror loss and loss due to the gain medium respectively. Loss due to gain medium will be calculated by integrating the imaginary part of metal permittivity in the desired frequency along the path of SPPs and loss due to mirrors will be calculated by Fresnel's law [22]. Using Lumerical FDTD simulations "γ" is calculated to be 6.2896×10$^3$ cm$^{-1}$.

In order to compare performance of plasmon lasers, there are various figures of merit. However, in this paper, we will use the threshold pump rate, Purcell factor, β factor, output power and operational bandwidth. Output power as can be witnessed in (8) is a function of the number of generated plasmons per unit volume of the cavity and can be derived from the rate equations of (5). [6]

$$P_{out} = \frac{1}{2} \times \frac{\alpha_m}{\alpha_m + \alpha_i} \times \frac{S}{\tau_p} \times \frac{hc}{\lambda} \times V_{mode} \quad (8)$$

Where "α$_m$" and "α$_i$" are mirror loss and intrinsic cavity loss respectively, "S" is plasmon number per unit volume, "τ$_p$" is plasmon lifetime in the cavity and equals to "Q/2π f$_{res}$" ("Q" is the quality factor and "f$_{res}$" is the resonant frequency of the cavity), "h" is Planck's constant, "c" is light speed, "λ" is the output wavelength and "V$_{mode}$" is mode volume. The bandwidth of the proposed nanolaser is characterized by two main time constants as can be seen in (9).

$$BW = \frac{1}{2\pi} \left( \frac{1}{\tau_{elec}} + \frac{1}{\tau_{plasmon}} \right) \quad (9)$$

The first parameter is electronic delay between input switching and change in carrier generation rate "τ$_{elec}$", which is determined by a parasitic RC time constant of the resonator and transit time of the carriers across the cavity, which can be calculated by (10):

$$\tau_{elec}^{-1} = \left( R_s \varepsilon_c \varepsilon_0 \frac{W_R^2}{H_R} \right)^{-1} + \tau_{resonator}^{-1}$$
$$\tau_{elec} \simeq \left( R_s \varepsilon_c \varepsilon_0 \frac{W_R^2}{H_R} \right) \quad (10)$$

Where "R$_s$" is internal resistance of electrical source and "ε$_c$" is average dielectric constant of both Germanium QWs and SiGe barriers and "τ$_{resonator}$" is transit time across the resonator consists of lifetime in QWs and transit time in the barriers. Nevertheless, it is about three orders of magnitude smaller than RC transit time and negligible in our case. Thus, "τ$_{elec}$" is equal to "6.5 ps" for our device.

The second parameter is "τ$_{plasmon}$" which contributes for SPP dynamics that can be calculated from 3dB bandwidth of spectral response transfer function of (11). [6]

$$\tau_{plasmon} = \frac{1}{\omega_{3dB}}$$
$$\omega_{3dB} : \omega @ H(\omega) = \frac{1}{2} H(0) \quad (11)$$
$$H(\omega) = \frac{\beta A(1 + S_0)}{\sqrt{\left(\omega^2 - \omega_r^2\right)^2 + \omega^2 \omega_p^2}}$$

Where "ω$_r$" and "ω$_p$" are derived from (12) and (13) respectively and "S$_0$" is the steady-state plasmon number. [6]

$$\omega_r = \sqrt{A\left[\frac{1+\beta S_0}{\tau_p} - AN_0\beta(1-\beta)\right]} \qquad (12)$$

Where "$N_0$" is steady state population inversion number. [5]

$$\omega_p = \frac{1}{\tau_p} + A(1 - \beta N_0 + \beta S_0) \qquad (13)$$

## 4. Results and output characteristics

One of the most important characteristics of a laser is output power profile vs normalized pump rate ($P/P_{th}$) which is shown in Fig.9. The behavior of this profile demonstrates the proper laser operation of the introduced structure.

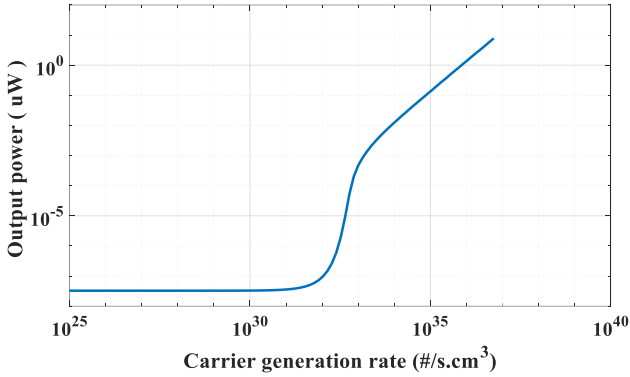

**Fig.9** Output power (μW) vs normalized pump rate

In addition, in Fig.10 a better input-output characteristic that relates the output SPP power (μW) to the input injection current (μA) is shown. Relatively large output power levels while maintaining the input pump current in microampere levels and in the room temperature result in a practically appropriate device for integration processes. In order to prove it, a thermal analysis using "Lumerical Device tool" [23] was performed. Thanks to the metallic waveguide structure, temperature distribution of the device in the relatively large pump current densities is appropriate for performing at room temperature without thermal breakdown.

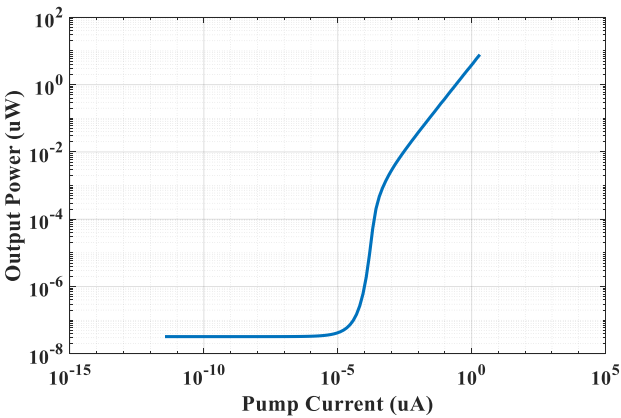

**Fig.10** Output power (μW) vs injected current (μA)

Finally, in Table.2 the key parameters of the proposed nanolaser are concluded. Although analysis done in this paper are based on theoretical models, they cannot guarantee if implemented it should work up to the derived performance. However, as mentioned before notable improvements over its competitors can be predicted.

**Table.2** Key parameters of the nanolaser

| Parameter | Value |
| --- | --- |
| Area ( square μm ) | 0.07 |
| Threshold current (pA) | 10 |
| Output power in μW ( pump = 1 μA ) | 3.83 |
| Modulation Bandwidth (GHz) | 24.5 |
| Purcell factor ( Lasing mode ) | 2208 |
| Coupling factor ( β ) | 0.324 |

## 5. Conclusion

In this paper a Ge/SiGe multi quantum well plasmon source was introduced, theoretically analyzed, and numerically simulated. The key advantages of the proposed structure are its tiny footprint (0.07μm$^2$), CMOS compatible process, room temperature operation, electrically pumping and high-efficiency coupling with metal/insulator plasmonic waveguides, which makes it a proper choice for the plasmon source in the development of plasmonic integrated circuits. The new structure generates 3.83μW output power with 1μA injection current in 1550nm free space frequency, has a wide modulation frequency of 24.5GHz, large Purcell factor about 2208 and very high output coupling ratio to the host plasmonic waveguide.